\documentclass[11pt]{article}
\usepackage{hyperref}
\usepackage{latexsym}
\usepackage{amssymb}
\usepackage{epsfig}
\usepackage{rotating}
\usepackage{amsfonts}
\usepackage{amsmath}
\usepackage{graphicx}
\newcommand\pngright[4]{
\Zw=#2 \divide \Zw by 5
\Zh=#3 \divide \Zh by 60  \advance\Zh by 1
\setcounter{wrapwidth}{\Zw}
\begin{wrapfigure}[\Zh]{r}{\value{wrapwidth}pt}
\begin{center}
\vspace{#4pt}
\includegraphics*[width=\Zw pt]{images/#1}
\end{center}
\end{wrapfigure}}

\newlength{\Taille}

\input xy
\xyoption{all}

\newcommand{\flechebas}[1]{
  \settoheight{\unitlength}{\mbox{$#1$}}
  \settowidth{\Taille}{\mbox{~${\scriptstyle #1}$}}
  \addtolength{\unitlength}{4ex}
  \begin{picture}(0,1)
    \put(0,1){\vector(0,-1){1}}
    \put(0,0.5){\makebox(0,0){${\scriptstyle #1}$ \hspace{\the\Taille}}}
  \end{picture}}
\newcommand{\flechehaut}[1]{
  \settoheight{\unitlength}{\mbox{$#1$}}
  \settowidth{\Taille}{\mbox{~${\scriptstyle #1}$}}
  \addtolength{\unitlength}{4ex}
  \begin{picture}(0,1)
    \put(0,0){\vector(0,1){1}}
    \put(0,0.5){\makebox(0,0){\hspace{\the\Taille}${\scriptstyle #1}$ }}
  \end{picture}}
\newcommand{\flechedroite}[1]{
  \settowidth{\unitlength}{\mbox{$#1$}}
  \settoheight{\Taille}{\mbox{${\scriptstyle #1}$}}
  \addtolength{\Taille}{1ex}
  \addtolength{\unitlength}{4ex}
  \raisebox{0.5ex}{
  \begin{picture}(1,0)
    \put(0,0){\vector(1,0){1}}
    \put(0.5,0){\makebox(0,0){${\scriptstyle #1}$ \vspace{\the\Taille}}}
  \end{picture}}}
\newcommand{\flechegauche}[1]{
  \settowidth{\unitlength}{\mbox{$#1$}}
  \settoheight{\Taille}{\mbox{${\scriptstyle #1}$}}
  \addtolength{\Taille}{1ex}
  \addtolength{\unitlength}{4ex}
  \raisebox{0.5ex}{
  \begin{picture}(1,0)
    \put(1,0){\vector(-1,0){1}}
    \put(0.5,0){\makebox(0,0){${\scriptstyle #1}$ \vspace{\the\Taille}}}
  \end{picture}}}

\newtheorem{definition}{Definition}[section]

\newtheorem{proposition}{Proposition}[section]

\newtheorem{remark}{Remark}[section]

\begin{document}
\begin{titlepage}
\begin{flushright}
{\it \bf ICMPA-MPA/2012}
\end{flushright}
\begin{center}
{\LARGE \bf {Twisted
Yang-Mills field theory: connections and Noether
current}}

Mahouton Norbert Hounkonnou$^{1,\dag}$ and Dine Ousmane
Samary$^{1,*}$

 $^{1}${\em University of Abomey-Calavi,\\
International Chair in Mathematical Physics
and Applications}\\
{\em (ICMPA--UNESCO Chair), 072 B.P. 50  Cotonou, Republic of Benin}\\

E-mails:   $^{\dag}$norbert.hounkonnou@cipma.uac.bj,\\
$^{*}$ousmanesamarydine@yahoo.fr.

\begin{abstract}
Main  properties of noncommutative (NC) gauge theory  are investigated in a  $2-$dimensional twisted Moyal
plane,  generated by  vector fields
$X_{a}=e_{a}^{\mu}(x)\partial_{\mu};$  the dynamical effects are
induced  by a non trivial  tensor $e_{a}^{\mu}(x)$. 
Connections in such  a NC space are defined. Symmetry analysis   is performed and related NC action is proved 
to be invariant under defined NC gauge transformations. 
A locally conserved Noether current  is explicitly computed.  Both commuting and noncommutative vector
 fields $X_{a}$ are considered.
\end{abstract}


\end{center}
\end{titlepage}

\section{Introduction}
The construction of noncommutative field theories in a nontrivial
background metric generally  implies a
non-constant deformation matrix
$\widetilde{\Theta}^{\mu\nu}=\widetilde{\Theta}^{\mu\nu}(x).$
There naturally results the  difficulty of finding a suitable explicit closed  Moyal type formula
and consequently, defining a noncommutative product becomes rather
complicated.The situation is simpler when one deals with the Moyal space
 $\mathbb{R}_{\Theta}^{D},$ i.e. the
deformed $D-$dimensional space endowed with a constant Moyal
$\star-$bracket of coordinate functions $[x^\mu,
x^{\nu}]_{\star}=i\Theta^{\mu\nu}$. In this case the star product (see
\cite{seiberg}-\cite{Das} and reference therein) is defined by
\begin{eqnarray}
(f\star g)(x) =  m\Big \{e^{i \frac{\Theta^{\mu\nu}}{2}
 \partial_\mu\otimes \partial_\nu} f(x)\otimes g(x)
 \Big \} \quad x\in \mathbb{R}_\Theta^D \quad  \forall f,g\in C^{\infty}(\mathbb{R}_\Theta^D)
\end{eqnarray}
$ m$ is the   ordinary multiplication of functions, i.e.
$m(f\otimes g)=f.g$.
 In the coordinate basis, this space is generated by the usual
commuting vector
 field $\partial_{\mu}=:\frac{\partial}{\partial x^\mu}\in T_x\mathbb{R}_{\Theta}^{D},$
 the tangent space of $\mathbb{R}_{\Theta}^{D}$,
  conferring  to Moyal
  space the properties of a flat space. 

On the contrary, in the  context of a dynamical noncommutative
field theory, the vector field can be generalized to take the
form $X_{a}=e_{a}^{\mu}(x)\partial_{\mu}$, where $e_{a}^{\mu}(x)$
is a tensor depending on the coordinate functions in the complex general  linear  matrix
group of order $D,$
GL$(D,\mathbb{C})$. The star product then takes the form
\begin{eqnarray}\label{pppp}
 (f\star g)(x) =  m\Big \{e^{i \frac{\Theta^{ab}}{2}
 X_{a}\otimes X_{b}} f(x)\otimes g(x)
 \Big \} \quad x\in \mathbb{R}_{\tilde\Theta}^D \quad  \forall f,g\in C^{\infty}(\mathbb{R}_{\tilde\Theta}^D)
 \end{eqnarray}
 and the vielbeins are given by the infinitesimal affine
transformation  as
\begin{eqnarray}\label{vie}
e_{a}^{\mu}(x)=\delta_{a}^{\mu}+\omega_{ab}^{\mu}x^b,
\end{eqnarray}
 where $\omega_{ab}\in \;$GL$(D,\mathbb{C})$. Using (\ref{vie}), the  non vanishing
Lie bracket peculiar to  the non-coordinate base   \cite{hd2}
\begin{eqnarray}
[X_a,X_b]=e_\nu^c\Big[e_a^\mu\partial_\mu
e_b^\nu-e_b^\mu\partial_\mu e_a^\nu\Big] X_c= C_{ab}^c X_c
\end{eqnarray}
 is here simply reduced to
\begin{eqnarray}\label{rela}
[X_a,X_b]=\omega_{ba}^\mu\partial_{\mu}-\omega_{ab}^\mu\partial_{\mu}=-2\omega_{ab}^\mu\partial_{\mu}.
\end{eqnarray}
Besides, the dynamical star product  (\ref{pppp}) can be now expressed as
\begin{eqnarray}\label{prod}
(f\star g)(x)=m\Big[\exp\Big(\frac{i}{2}\theta
e^{-1}\epsilon^{\mu\nu}\partial_\mu\otimes\partial_\nu\Big)(f\otimes
g)(x)\Big] 
\end{eqnarray}
where  $e^{-1}=:det(e_{a}^{\mu})=1+\omega_{12}^{1}x^{2}-\omega_{12}^{2}x^{1}$;\,
$\epsilon^{\mu\nu}$ is the symplectic tensor in two dimensions, $(D=2),$ i.e
$\epsilon^{12}=-\epsilon^{21}=1,\,\,
\epsilon^{11}=\epsilon^{22}=0$.

The coordinate function commutation relation becomes
 $[x^\mu, x^{\nu}]_{\star}=i\widetilde{\Theta}^{\mu\nu}=i(\Theta^{\mu\nu}-\Theta^{a[\mu}\omega_{ab}^{\nu]}x^{b})$
 which can be reduced to the usual Moyal space relation, as expected, by setting
 $\omega_{ab}^{\mu}=[0]$.
One can  check that the Jacobi identity is also well satisfied, i.e. 
 \begin{eqnarray}\label{id}
 [x^\mu,[x^\nu, x^{\rho}]_{\star}]_{\star}+[x^\rho,[x^\mu, x^\nu]_{\star}]_{\star}
 +[x^\nu,[x^\rho, x^\mu]_{\star}]_{\star}=\Theta^{b\mu}\Theta^{d[\nu}\omega_{bd}^{\rho]}=0
\end{eqnarray}
conferring a Lie algebra structure to the defined twisted Moyal
space.
This identity   ensures the  associativity of the
star-product (\ref{pppp}) and implies that
\begin{eqnarray}
 \widetilde{\Theta}^{\sigma\rho}\partial_{\rho}\widetilde{\Theta}^{\mu\nu}
 +\widetilde{\Theta}^{\nu\rho}\partial_{\rho}\widetilde{\Theta}^{\sigma\mu}
 +\widetilde{\Theta}^{\mu\rho}\partial_{\rho}\widetilde{\Theta}^{\nu\sigma}=0.
\end{eqnarray}
 Remark that with the relation
(\ref{rela}), the requirement that $\omega_{ab}$ is a symmetric
tensor trivially ensures  the associativity of the star product. 
In the interesting particular case addressed in this work,  the associativity 
of the star product (\ref{pppp}) can be shown even for the non symmetric 
tensor $\omega_{ab}$. See  appendix for details.

From the particular condition $[X_{a},X_{b}]=0,$ (i. e. the vector
fields are commuting), there result
  constraints on $e_{a}^{\mu}$, namely $e_{[a}^{\mu}\partial_{\mu}e_{b]}^{\nu}=0$,
  that can be solved off-shell in terms of $D$ scalar fields $\phi^{a}$.
   Supposing that the square matrix $e_{a}^{\mu}$ has an inverse $e_{\mu}^{a}$ everywhere so that
   the  $X_{a}$ are linearly independent,
    then the above condition becomes $\partial_{[\mu}e_{\nu]}^{a}=0$ which is satisfied by $e_{\nu}^{a}=\partial_{\nu}\phi^{a}$.
Since $X_{a} \phi^b = \delta_a^b$, the field $\phi^b$ can be viewed
as new coordinates along the $X_a$ directions.  The metric {\rm g}
on
 $\mathbb{R}_{\tilde\Theta}^{D}$ can be chosen to be $\mbox{\rm g}(X_a,X_b)=e_a^\mu e_b^\nu
 \mbox{\rm g}_{\mu\nu}=\delta_{ab}$. See
\cite{aschieri}-\cite{ hd} for more details. In the whole work, we deal with Euclidean signature and  $D=2$.

The paper is organized as follows. In Section 2, we provide
 the general properties of gauge theory in twisted
Moyal space, and define related connections.  The tensor $\omega_{ab}^\mu$ is an
 infinitesimal tensor, skew-symmetric in the indexes $a$ and $b$. In Section 3, we study the symmetry of
pure gauge theory and  show that the related NC action is
invariant under  $U_\star(1)$ gauge transformation. Besides, we compute the resulting Noether current. In Section 4, we
investigate the properties of the model for  commuting
vector fields $X_a$.  Section 5 is devoted to  concluding remarks.

\section{Connections and gauge transformation}
Consider $E=\{x^\mu, \mu\in [[1,2]]\}$ and
$\mathbb{C}[[x^{1}, x^{2}]],$ the free algebra generated by $E$.
Let $\mathcal{I}$ be the ideal of $\mathbb{C}[[x^{1}, x^{2}]],$
engendered by the elements $x^\mu\star x^\nu-x^\nu\star x^\mu
-i\tilde{\Theta}^{\mu\nu}$. The twisted Moyal Algebra
$\mathcal{A}_{\tilde\Theta}$ is the quotient $\mathbb{C}[[x^{1},
x^{2}]]/\mathcal{I}$. Each element in $\mathcal{A}_{\tilde\Theta}$ is a
formal power series in the $x^\mu$'s for which the relation
$[x^\mu, x^\nu]_\star=i\tilde{\Theta}^{\mu\nu}$ holds. Moyal
algebra can be here also defined as  the linear space of smooth and rapidly decreasing
functions equipped with the  NC star product given in
(\ref{pppp}).
 The gauge symmetries
on this noncommutative space can be realized in their enveloping
algebra. However, there is an isomorphism mapping  the
noncommutative functions algebra $\mathcal{A}_{\tilde\Theta}$ into the
commutative one, equipped with an additional noncommutative
$\star-$product \cite{seiberg}.
We consider the following infinitesimal affine transformation
\begin{eqnarray}
e_{a}^{\mu}(x)=\delta_{a}^{\mu}+\omega_{ab}^{\mu}x^{b},\quad
\omega_{ab}^{\mu}= :-\omega_{b a}^{\mu},\mbox{ and }
|\omega^{\mu}|<<1.
\end{eqnarray}
For $D=2$,
 $e_{a}^{\mu}$ and $\Theta^{ab}$ can be expressed as follows:
\begin{eqnarray}
 (e)_{a}^{\mu}=\Big(\begin{array}{cc}
1+\omega_{12}^{1}x^{2}&\omega_{12}^{2}x^{2}\\
-\omega_{12}^{1}x^{1}&1-\omega_{12}^{2}x^{1}
\end{array}\Big) \quad \mbox{ and }\quad (\Theta)^{ab}=\Big(\begin{array}{cc}
0&\theta\\
-\theta&0
\end{array}\Big)=\theta (\epsilon)^{ab}
 \end{eqnarray}
where $\epsilon^{12}=-\epsilon^{21}=1,\,\,
\epsilon^{11}=\epsilon^{22}=0$. There follow the relations
\begin{eqnarray}
e^{-1}=:det(e_{a}^{\mu})=1+\omega_{12}^{1}x^{2}-\omega_{12}^{2}x^{1}\\
 e=:det(e_{\mu}^{a})= 1-\omega_{12}^{1}x^{2}+\omega_{12}^{2}x^{1}.
\end{eqnarray}
The tensor  $\widetilde{\Theta}^{\mu\nu}$ can be  written  as
\cite{hd2}
\begin{eqnarray}
(\widetilde{\Theta})^{\mu\nu}=(\Theta)^{\mu\nu}-(\Theta^{a[\mu
}\omega_{ab}^{\nu]})x^b =\Big(\begin{array}{cc}
0& \theta e^{-1}\\
-\theta e^{-1}&0
\end{array}\Big).
\end{eqnarray}
Let us now define the space-time
($M\subseteq\mathbb{R}_{\tilde\Theta}^{2}$) metric as
\begin{eqnarray}
\mbox{\rm g}_{\mu\nu}=e_{\mu}^{a}e_{\nu}^{b}\delta_{ab}=\Big(
\begin{array}{cc}
1-2\omega_{12}^{1}x^{2}&\omega_{12}^{1}x^{1}-\omega_{12}^2x^2\\
\omega_{12}^{1}x^{1}-\omega_{12}^2x^2&1+2\omega_{12}^{2}x^1
\end{array}
\Big),\,\,\\
\mbox{ where }\,\, e_{\mu}^{a}=\Big(
\begin{array}{cc}
1-\omega_{12}^{1}x^{2}&-\omega_{12}^{2}x^2\\
\omega_{12}^{1}x^1&1+\omega_{12}^{2}x^1
\end{array}\Big)
\end{eqnarray}
and its inverse as
\begin{eqnarray}
\mbox{\rm g}^{\mu\nu}=e^{\mu}_{a}e^{\nu}_{b}\delta^{ab}=\Big(
\begin{array}{cc}
1+2\omega_{12}^{1}x^{2}&\omega_{12}^{2}x^{2}-\omega_{12}^{1}x^1\\
\omega_{12}^{2}x^{2}-\omega_{12}^{1}x^1&1-2\omega_{12}^{2}x^1
\end{array}
\Big),\,\,\\
\mbox{ where }\,\, e^{\mu}_{a}=\Big(
\begin{array}{cc}
1+\omega_{12}^{1}x^{2}&\omega_{12}^{2}x^2\\
-\omega_{12}^{1}x^1&1-\omega_{12}^{2}x^1
\end{array}\Big)
\end{eqnarray}
with ${\rm g}=-det(\mbox{\rm g}_{\mu\nu})$. Noncommutative field
theory over Moyal algebra of functions can be defined as field
theories over module $\mathcal{H}$ on the noncommutative algebra
$\mathcal{A}_{\tilde\Theta}$ or as matrix theories with coefficients in
$\mathcal{A}_{\tilde\Theta}$. In the following, we restrict the study of
field modules to rank trivial bi-modules $\mathcal{H}$ over
$\mathcal{A}_{\tilde\Theta}$ with a Hilbert space structure defined by
the scalar product
\begin{eqnarray}
<a,b>=:\int\,\,ed^2x\,\,Tr(a^{\dag}\star b)\star e^{-1}; \,\,\,
a,b\in \mathcal{A}_{\tilde\Theta}.
\end{eqnarray}
Provided this framework, the notion of
connection defined on vector bundles in ordinary differential
geometry is replaced, in NC geometry, by the generalized concept
of connection on the projective modules as follows.
\begin{definition}
 The
sesquilinear maps $\nabla_\mu: \mathcal{H}\rightarrow\mathcal{H} $
are called connections if they satisfy the differentiation chain
rule
\begin{eqnarray}
\nabla_\mu(m\star f )=m\star
\partial_{\mu}(f)+\nabla_\mu(m)\star f,\,\,\mbox{ for } f\in \mathcal{A}_{\tilde\Theta}
\mbox{ and } m\in \mathcal{H}
\end{eqnarray}
 (assumed here to be a
right module over $\mathcal{A}_{\tilde\Theta}$), and if they are
compatible with the Hermitian structure of $\mathcal{H}$ defined
as ${\rm h}(f,g)=f^\dag\star g$, i.e.
\begin{eqnarray}
\partial_{\mu}{\rm h}(m_1,m_2)={\rm h}(\nabla_{\mu}m_1,m_2)+{\rm
h}(m_1,\nabla_{\mu}m_2).
\end{eqnarray}
\end{definition}
In the sequel,  we can identify $\mathcal{A}_{\tilde\Theta}$ with
$\mathcal{H}$.
\begin{definition}
Denoting by ${\bf 1}$ the unit element of $\mathcal{A}_{\tilde\Theta}$,
we define the gauge potential by $\nabla_\mu {\bf 1} =-iA_\mu$.
Then the connection can be explicitly written as
\begin{eqnarray}
\nabla_\mu(.)=\partial_\mu(.)-iA_\mu\star(.)
\end{eqnarray}
$A_\mu$ is called the gauge potential in the fundamental
representation.
\end{definition}
Note that the left module can be used to define the connection in
the anti-fundamental representation by
$\nabla_\mu(.)=\partial_\mu(.)+i(.)\star A_\mu$. In the same vein,
the module can be used to define the connection on the adjoint
representation by
$\nabla_\mu(.)=\partial_\mu(.)-i[A_\mu,(.)]_\star$. Here, we adopt
 the fundamental representation. Now, we  define the
gauge transformation as a morphism of module, denoted by $\gamma$,
 satisfying the relation
\begin{eqnarray}
\gamma(m\star f)=\gamma(m)\star f\,\,\mbox{ for } f\in
\mathcal{A}_{\tilde\Theta} \mbox{ and } m\in \mathcal{H}
\end{eqnarray}
and preserving the Hermitian structure ${\rm h}$, i.e.
\begin{eqnarray}
{\rm h}(\gamma(f),\gamma(g))={\rm h}(f,g)\,\, \mbox{ for } f,g\in
\mathcal{A}_{\tilde\Theta}.
\end{eqnarray}
For $f=g=1$, one can prove that $\gamma(1)\in
U(\mathcal{A}_{\tilde\Theta}),$ the group of unitary elements of
$\mathcal{A}_{\tilde\Theta},$ i.e. $\gamma^{\dag}(1)\star \gamma(1)=1$.

Note that the Jacobi identity is covariantly written  in the form
\begin{eqnarray}
 \widetilde{\Theta}^{\sigma\rho}\nabla_{\rho}\widetilde{\Theta}^{\mu\nu}
 +\widetilde{\Theta}^{\nu\rho}\nabla_{\rho}\widetilde{\Theta}^{\sigma\mu}
 +\widetilde{\Theta}^{\mu\rho}\nabla_{\rho}\widetilde{\Theta}^{\nu\sigma}=0.
\end{eqnarray}
This equation is evidently satisfied whenever the following
condition holds
\begin{eqnarray}
\nabla_{\rho}\widetilde{\Theta}^{\mu\nu}=0,
\end{eqnarray}
which is very simple to handle in two-dimensional space. Indeed, in $D=2$
the most general $\widetilde{\Theta}$ can be written  in
the form
\begin{eqnarray}
\widetilde{\Theta}^{\mu\nu}=\frac{\epsilon^{\mu\nu}}{\sqrt{-\mbox{\rm
g}(x)}}\theta(x^1, x^2),
\end{eqnarray}
where $\theta(x^1, x^2)$ is a constant, simply denoted by
$\theta$. Then
\begin{eqnarray}
e^{-1}=1/\sqrt{-\mbox{\rm g}(x)}\,\,\,\mbox{ or } \,\,\,
e=\sqrt{-\mbox{\rm g}(x)}.
\end{eqnarray}

To end this section, let us mention that the integral $\int\,d^Dx\, f\star g$, defined with the dynamical
Moyal $\star-$product (\ref{pppp}),  is not cyclic, even with
suitable boundary conditions at infinity, i.e.
\begin{eqnarray}
\int\,\mbox{d}^{D}x \, (f\star g)\neq \int\,\mbox{d}^{D}x \,
(g\star f).
\end{eqnarray}
Using now the measure $ed^Dx$ where $e=det(e_\mu^a)$, a cyclic
integral can be defined so that, up to boundary terms:
\begin{eqnarray}
 \int\,e\mbox{d}^{D}x \, (f\star g)=\int\,e\mbox{d}^{D}x(fg)=\int\,e\mbox{d}^{D}x \, (g\star f).
\end{eqnarray}
In flat space $\sqrt{-\mbox{\rm g}(x)}=1$. 

\section{Dynamical pure gauge theory}
We consider a field $\psi,$  element of the algebra
$\mathcal{A}_{\tilde\Theta}$, ($\psi\in\mathcal{A}_{\tilde\Theta}$), and
its infinitesimal gauge variation
$\delta \psi$  such that, under an
infinitesimal gauge transformation $\alpha(x)$, the relation  $\delta_{\alpha}\psi(x)=i\alpha(x)\star\psi(x)$ is obeyed. The
covariant coordinates are defined as
\begin{eqnarray}
X^{\mu}=x^{\mu}+A^{\mu}, \quad A^{\mu}\in\mathcal{A}_{\tilde\Theta}
\end{eqnarray}
$A^{\mu}$ is called the gauge potential and satisfies the relation
$\delta_{\alpha}A^{\mu}=\tilde{\Theta}^{\mu\rho}\partial_{\rho}
\alpha+i[\alpha,A^{\mu}]_{\star}$. One can check that
\begin{eqnarray}
[\alpha(x),\tilde{\Theta}^{\mu\sigma}]_{\star}=i\Theta^{a[\mu}
\omega_{ac}^{\sigma]}\Theta^{c\rho}\partial_{\rho}\alpha(x),\,
\,\mbox{ and }\,\, \tilde{\Theta}_{\mu\sigma}^{-1}\delta_{\alpha}
\tilde{\Theta}^{\mu\sigma}=2\omega_{ac}^{a}\delta x^{c}.
\end{eqnarray}
From the last two equations and the definition of
 $A_{\sigma}$ such that
$A^{\mu}=\tilde{\Theta}^{\mu\sigma}A_{\sigma},$  we derive the gauge variation
\begin{eqnarray}\label{ga}
\delta_{\alpha}A_{\sigma}=\partial_{\sigma}\alpha(x)+i[\alpha(x)
,A_{\sigma}]_{\star}+2\omega_{ac}^{a}\Big(\Theta^{c\rho}\partial_{\rho}\alpha(x)-\delta
x^{c}\Big)A_{\sigma}
\end{eqnarray}
There result    two contributions in the expression of $\delta_{\alpha}A_{\sigma}$: the first contribution consisting of the
first two terms of the ordinary Moyal product \cite{madore}, and
the second one given by the last term pertaining to the
twisted effects of the theory. Of course, when $\omega=0$, we
recover the usual Moyal result. 

The NC gauge
tensor $T^{\mu\nu}\in\mathcal{A}_{\tilde\Theta}$ is defined by
$T^{\mu\nu}=[X^{\mu}, X^{\nu}]_{\star}-i\tilde{\Theta}^{\mu\nu}$
and satisfies the properties
\begin{eqnarray}\label{ga2}
\delta_{\alpha}T^{\mu\nu}=i[\alpha(x),T^{\mu\nu}]_{\star}.
\end{eqnarray}
It is then convenient to use   the relation
$T^{\mu\nu}=i\tilde{\Theta}^{\mu\sigma}\tilde{\Theta}^{\nu\lambda}F_{\sigma\lambda}$
to derive the twisted Faraday tensor $F_{\sigma\lambda}$ as 
\begin{eqnarray} 
F_{\sigma\lambda}&=&
\partial_{\sigma}A_{\lambda}-\partial_{\lambda}A_{\sigma}-i[A_{\sigma}
,A_{\lambda}]_{\star}-\Theta_{\nu\lambda}^{-1}\Theta^{a[\nu}\omega_{a\sigma}^{\lambda]}A_\lambda
+\Theta_{\mu\sigma}^{-1}\Theta^{a[\mu}\omega_{a\lambda}^{\sigma]}A_\sigma\cr
&&-\Big(\Theta_{\mu\sigma}^{-1}\Theta^{a[\mu}\omega_{ac}^{\sigma]}\Theta^{c\rho}\partial_{\rho}A_\lambda\Big)A_\sigma
+\Big(\Theta_{\nu\lambda}^{-1}\Theta^{a[\nu}\omega_{ac}^{\lambda]}\Theta^{c\rho}\partial_{\rho}A_{\sigma}\Big)A_\lambda\\
&=&\partial_{\sigma}A_{\lambda}-\partial_{\lambda}A_{\sigma}-i[A_{\sigma}
,A_{\lambda}]_{\star}+2\omega_{a[\sigma}^{a}A_{\lambda]}-2\omega_{ac}^{a}
\Theta^{c\rho}(\partial_{\rho}A_{[\sigma})A_{\lambda]}.
\end{eqnarray}
Canceling the "{\it twisted}" contributions involved in the last
two terms on the right hand side of this relations, we turn back
to  usual Moyal product result.

We then arrive at the expression of the dynamical NC
pure gauge action  defined as
\begin{eqnarray}\label{ac}
\mathcal{S}_{YM}=-\frac{1}{4\kappa^2}\int\,\, ed^{2}x\,\,
\Big(F_{\mu\nu}\star F^{\mu\nu}\star e^{-1}\Big)
\end{eqnarray}
where $e=: det(e_{\mu}^{a})$.

Now, expanding
 the dynamical $\star$-product  (\ref{pppp}) of two functions
  as follows:
\begin{eqnarray}\label{starpro}
f\star g&=&fg+\frac{i}{2}\Theta^{ab}X_{a}fX_{b}g\cr
&&+\frac{1}{2!}\Big(\frac{i}{2}\Big)^{2}
\Theta^{a_{1}b_{1}}\Theta^{a_{2}b_{2}}(X_{a_{1}}X_{a_{2}}f)(X_{b_{1}}X_{b_{2}}g) +\cdots\nonumber\\
&\equiv & e^{\Delta}(f,g)
\end{eqnarray}
where powers of the bilinear operator $\Delta$ are defined as
\begin{eqnarray}
\Delta(f,g)=\frac{i}{2}\Theta^{ab}(X_{a}f)(X_{b}g)\qquad \Delta^{0}(f,g)=fg\nonumber\\
\Delta^{n}(f,g)=\Big(\frac{i}{2}\Big)^{n}\Theta^{a_{1}b_{1}}\cdots\Theta^{a_{n}b_{n}}(X_{a_{1}}
\cdots X_{a_{n}}f)(X_{b_{1}}\cdots X_{b_{n}}g)
\end{eqnarray}
 one can deduce the following rules
 (straightforwardly generalizing  the  usual Moyal product identities):
\begin{eqnarray}
&f\star g=fg+X_{a}T(\Delta)(f,\widetilde{X}^{a}g)\\
&[f,g]_{\star}=f\star g-g\star f =2X_{a}S(\Delta)(f,\widetilde{X}^{a}g)\\
&\{f,g\}_{\star}=f\star g+ g\star
f=2fg+2X_{a}R(\Delta)(f,\widetilde{X}^{a}g)
\end{eqnarray}
with
\begin{eqnarray}
T(\Delta)&=&\frac{e^{\Delta}-1}{\Delta}\qquad
S(\Delta)=\frac{sinh(\Delta)}{\Delta}\nonumber
\\ R(\Delta)&=&\frac{cosh(\Delta)-1}{\Delta}\mbox{ and } \widetilde{X}^{a}=\frac{i}{2}\Theta^{ab}X_{b}.
\end{eqnarray}
 $S(\Delta)(.,\widetilde{X}.)$ is a bilinear antisymmetric operator and
 $T(\Delta)(f,\widetilde{X}^{a}g)-T(\Delta)(g,\widetilde{X}^{a}f)=2S(\Delta)(f,\widetilde{X}^{a}g).$

Define also the gauge transformation $U$ by
\begin{eqnarray}\label{tran}
 U&=&e^{i\alpha}_{\star}=\sum_{k=0}^{\infty}\frac{i^{k}(\alpha)_\star^k}{k!}\cr
 &=&1+i\alpha+
 (i^{2}/2!)\alpha\star\alpha+(i^{3}/3!)\alpha\star\alpha\star\alpha+\cdots;\,\,\, \alpha\in
 C^{\infty}(\mathbb{R})
\end{eqnarray}
$U_{\star}(1)$ is NC gauge group generated by elements $U\in
U_{\star}(1)$. The infinitesimal gauge transformation
$U=1+i\alpha(x)$ defined in the noncommutative Moyal space is the
same as the ordinary infinitesimal gauge transformation in
commutative space. Making the gauge transformation of tensor $F_{\mu\nu}$
into $F_{\mu\nu}^U=U \star F_{\mu\nu}\star U^\dag$,  then the transformed gauge action
 yields:
\begin{eqnarray}
\mathcal{S}_{YM}^U&=&-\frac{1}{4\kappa^2}\int\,\, ed^{2}x\,\,
\Big(U\star F_{\mu\nu}\star U^\dag\star U\star F^{\mu\nu}\star
U^\dag \star e^{-1}\Big)\cr &=& -\frac{1}{4\kappa^2}\int\,\,
ed^{2}x\,\, \Big(U\star F_{\mu\nu}\star F^{\mu\nu}\star U^\dag
\star e^{-1}\Big)\cr &=& -\frac{1}{4\kappa^2}\int\,\, ed^{2}x\,\,
\Big( F_{\mu\nu}\star F^{\mu\nu}\star U ^\dag \star e^{-1}\star
U\cr &&+2S(\Delta)(U,\widetilde{X}^a(F_{\mu\nu}\star
F^{\mu\nu}\star U^\dag \star e^{-1}))\Big)
\end{eqnarray}
\begin{proposition}
Provided the stubborn requirement that the surface terms be vanished, the action
(\ref{ac}) is invariant  under the global gauge transformation, i.e. setting 
$\alpha= \alpha_0= c^{ste}$ in
(\ref{tran}).
\end{proposition}
{\bf Proof:}\,
 From
the infinitesimal gauge transformation $U(\alpha)=1+i\alpha(x)$, its
 conjugate  given by $U^\dag(\alpha)=1-i\alpha(x)$ and the definition
 $e^{-1}:=1+\omega_\mu x^\mu$, where
$\omega_1=\omega_{12}^1$ and $\omega_2=-\omega_{12}^2$, we have
\begin{eqnarray}
U^\dag\star
e^{-1}=e^{-1}(1-i\alpha(x))+\frac{1}{2}\widetilde{\Theta}^{\mu\nu}\partial_\mu\alpha(x)\omega_\nu\nonumber
 \end{eqnarray}
\begin{eqnarray}
\Rightarrow U^\dag \star e^{-1}\star
U&=&e^{-1}+\widetilde{\Theta}^{\mu\nu}\omega_\nu\partial_{\mu}\alpha(x)
=e^{-1}
+\Theta^{\mu\nu}\omega_{\nu}\partial_{\mu}\alpha(x).\nonumber
\end{eqnarray}
\begin{eqnarray}
&&U^\dag \star e^{-1}\star U=e^{-1}\Rightarrow
\Theta^{\mu\nu}\omega_{\nu}\partial_{\mu}\alpha(x)=0\Rightarrow
\alpha(x)=\alpha_0=c^{ste}\,\,\,\,\, \square\nonumber
\end{eqnarray}
Furthermore, the following statement is true.
\begin{proposition}
Provided the same requirement of vanishing condition of the surface terms, 
$(i)$ the action
(\ref{ac}) is invariant  under the noncommutative group of unitary
transformations $U_\star(1)$  defined
by the parameter $\alpha=\alpha_0 +\epsilon_\sigma x^\sigma$, where
$\epsilon_\sigma$ is an infinitesimal parameter and $\alpha_0$ an
arbitrary constant, and $(ii)$ there exists an isomorphism between the NC gauge group induced by (\ref{ga})
and    $U_\star(1)$ group.
\end{proposition}
{\bf Proof:}\,
The part $(i)$ is immediate from the previous proof.
$(ii)$  Imposing the condition  
$\alpha=\alpha_0+\epsilon_\lambda x^\lambda,$
 the NC gauge transformation (\ref{ga}) is reduced to the form:
\begin{eqnarray}
\delta_\alpha A_\sigma =\partial_\sigma\alpha-\epsilon_\mu\Theta^{\mu\rho}\partial_\rho A_\sigma
=\partial_\rho\Big(\delta_\sigma^\rho\frac{\alpha}{2}-\epsilon_\mu\Theta^{\mu\rho} A_\sigma\Big) =\partial \Lambda
\end{eqnarray}
giving rise to the isomorphism 
\begin{eqnarray}
f: \partial\Lambda\rightarrow e_\star^{i\Lambda}=1+i\Big(\frac{\alpha}{2}-\epsilon_\mu\Theta^{\mu\rho} A_\sigma\Big)
\end{eqnarray}
mapping the  NC gauge group (\ref{ga}) into the $U_\star(1)$ group. $\square$

Therefore,  the $U_\star(1)$ group can be  considered as the invariance 
 NC gauge group for the Yang-Mills action defined in (\ref{ac}). Note that setting $\epsilon_\mu=0$
we recover the  global gauge transformation of the usual  gauge field theory.

 The
$A_{\mu}$ variation of the action (\ref{ac}) is given by
\begin{eqnarray}
\delta_{A}\mathcal{S}_{YM}
&=&-\frac{1}{4\kappa^2}\int\,\,d^{2}x\,\,(\delta
A_{\beta}\mathcal{E}_{A}+\partial_{\beta} J^{\beta})
\end{eqnarray}
where the equation of motion of the field $A$ is provided by
\begin{eqnarray}
\frac{\delta \mathcal{S}_{YM}}{\delta
A_{\beta}}=\mathcal{E}_{A}&=&-\partial_{\mu}(e\{F^{\mu\beta},e^{-1}\}_{\star})+\partial_{\nu}(e\{F^{\beta\nu},e^{-1}\}_{\star})\cr
&&-ie[A_{\nu},\{F^{\beta\nu},e^{-1}\}_{\star}]_{\star}+ie[A_{\mu},\{F^{\mu\beta},e^{-1}\}_{\star}]_{\star}\cr
&&-2\omega_{ac}^{a}\Theta^{c\rho}\Big(\partial_{\rho}(-eA_{\nu}\{F^{\beta\nu},e^{-1}\}_{\star}
+eA_{\mu}\{F^{\mu\beta},e^{-1}\}_{\star})\Big)\cr
&&-2e\omega_{ac}^{a}\Theta^{c\rho}\Big(\partial_{\rho}A_{\mu}\{F^{\mu\beta},e^{-1}\}_{\star}-\partial_{\rho}A_{\nu}\{F^{\beta\nu},e^{-1}\}_{\star}\Big)\cr
&&+2e\Big(\omega_{a\mu}^{a}\{F^{\mu\beta},e^{-1}\}_{\star}-
\omega_{a\nu}^{a}\{F^{\beta\nu},e^{-1}\}_{\star}\Big) =0\nonumber\\
\end{eqnarray}
and the current $J^{\beta}$ by
\begin{eqnarray}\label{cure}
J^{\beta}&=&-\frac{1}{4\kappa^2}\Big[ e\delta
A_{\nu}\{F^{\beta\nu},e^{-1}\}_{\star}-e\delta
A_{\mu}\{F^{\mu\beta},e^{-1}\}_{\star}\cr
&&-iee_{a}^{\beta}\Big(T(\Delta)(\delta
A_{[\mu},\widetilde{X}^{a}[A_{\nu]},\{F^{\mu\nu},e^{-1}\}_{\star}]_{\star})\cr
&&+T(\Delta)([\delta
A_{[\mu},A_{\nu]}]_{\star},\widetilde{X}^{a}\{F^{\mu\nu},e^{-1}\}_{\star})\cr
&&+2S(\Delta)(A_{[\mu},\widetilde{X}^{a}(\delta
A_{\nu]}\{F^{\mu\nu},e^{-1}\}_{\star}))\Big)\cr
 &&-2\omega_{ac}^{a}\Theta^{c\beta}e\delta
A_{[\mu}A_{\nu]}\{F^{\mu\nu},e^{-1}\}_{\star}\cr
&&+ee_{a}^{\beta}\Big(T(\Delta)(\delta
F^{\mu\nu},\widetilde{X}^{a}\{F^{\mu\nu},e^{-1}\}_{\star})\cr
&&+2S(\Delta)(F^{\mu\nu},\widetilde{X}^{a}\delta F^{\mu\nu}\star
e^{-1})\Big)\Big]
\end{eqnarray}
where $[\delta A_{[\mu},A_{\nu]}]_{\star}=\delta A_{\mu}\star
A_{\nu}-A_{\nu}\star\delta A_{\mu}-\delta A_{\nu}\star
A_{\mu}+A_{\mu}\star\delta A_{\nu}  $. Using the property that
$F^{\mu\nu}=-F^{\nu\mu}$ and the fact that the surface terms are
canceled, the equation of motion $\mathcal{E}_{A}= 0$ and the
current $J^\beta$ can be re-expressed, respectively, as
\begin{eqnarray}\label{motion}
\frac{\delta \mathcal{S}_{YM}}{\delta
A_{\beta}}=\mathcal{E}_{A}&=&-2\partial_{\mu}(e\{F^{\mu\beta},e^{-1}\}_{\star})
-4\omega_{ac}^{a}\Theta^{c\rho}
\partial_{\rho}(eA_{\mu}\{F^{\mu\beta},e^{-1}\}_{\star})
\cr
&&-4e\omega_{ac}^{a}\Theta^{c\rho}\partial_{\rho}A_{\mu}\{F^{\mu\beta},e^{-1}\}_{\star}
+4e\omega_{a\mu}^{a}\{F^{\mu\beta},e^{-1}\}_{\star}=0
\end{eqnarray}
and
\begin{eqnarray}
J^\beta=\frac{1}{2\kappa^2}\Big(e\delta
A_{\mu}\{F^{\mu\beta},e^{-1}\}_{\star}+\omega_{ac}^{a}\Theta^{c\beta}e\delta
A_{[\mu}A_{\nu]}\{F^{\mu\nu},e^{-1}\}_{\star}\Big).
\end{eqnarray}
 Let us now deal with the symmetry analysis and deduce the
conserved currents. Performing the following functional variation of 
fields and  coordinate transformation
\begin{eqnarray}
A'_\mu(x)=A_\mu(x)+\delta A_\mu(x),\,\,\,\,\,\,\,\,
x'^{\mu}=x^{\mu}+\epsilon^{\mu},\quad \epsilon^{\mu}=\delta
x^\mu=0
\end{eqnarray}
and using
 $\mbox{d}^{2}x'=[1+\partial_{\mu}\epsilon^{\mu}+\textbf{O}(\epsilon^{2})]\mbox{d}^{2}x=\mbox{d}^{2}x$
lead to the following  variation of the action, to first order in
 $\delta A_\mu(x)$ and $\delta\phi^{c}(x)$:
\begin{eqnarray}
\delta\mathcal{S}_{YM}&=&
\int\,e\mbox{d}^{2}x\Big\{\Big{\vert}\frac{\partial x'}{\partial
x} \Big{\vert}\star(\mathcal{L'}_{YM}\star
e^{-1})\Big\}-\int\,e\mbox{d}^{2}x\,
 (\mathcal{L}_{YM}\star e^{-1})\nonumber\\
&=&\int\,\mbox{d}^{2}x\,\,\delta\Big((\mathcal{L}_{YM}\star
e^{-1})e\Big)  =\int{\mbox{d}}^2
x\hspace{1mm}\Big\{\delta_{A_\mu}\Big((\mathcal{L}_{YM}\star
e^{-1})e\Big)\Big)\Big\}
\end{eqnarray}
where
\begin{eqnarray}
\mathcal{L}_{YM}=-\frac{1}{4\kappa^2}F^{\mu\nu}\star
F_{\mu\nu}\,\,\mbox{ and }\,\,
\mathcal{L'}_{YM}=-\frac{1}{4\kappa^2}F^{\mu\nu}_U\star
F_{\mu\nu}^U.
\end{eqnarray}
 On shell, and integrated on a submanifold $M \subset \mathbb{
R}^2$ with fields non vanishing at the boundary (so that the total
derivative terms
 do not disappear),
 we get:
\begin{eqnarray}\label{noether}
\delta\mathcal{S}_{YM}= \int_{M}{\mbox{d}}^2 x\hspace{1mm}
\partial_{\sigma}\mathcal{J}^{\sigma}=0.
\end{eqnarray}
\begin{proposition}
The noncommutative Noether current $\mathcal{J}^{\sigma}$
 is locally conserved.
 \end{proposition}
 {\bf Proof:}\,
 The analysis of the local properties of this tensor
requires the useful formulas
\begin{eqnarray}
 \delta_{\alpha}
A_\mu=\epsilon_\mu(1+\Theta^{\rho\sigma}\partial_\rho
A_\sigma),\,\,\,\omega_{ac}^a=-\omega_c,\,\,\,
\partial_\beta e=-\omega_\beta,\,\,\,
\{F^{\mu\nu},e^{-1}\}_\star=2e^{-1}F^{\mu\nu}.
\end{eqnarray}
A straightforward computation gives
\begin{eqnarray}
  J^\beta=\frac{\epsilon_\mu}{\kappa^2}\Big(1+\Theta^{\rho\sigma}\partial_\rho
  A_\sigma\Big)F^{\mu\beta}\Rightarrow \partial_\beta J^\beta=\frac{\epsilon_\mu}{\kappa^2}\Big(1+\Theta^{\rho\sigma}\partial_\rho
  A_\sigma\Big)\partial_\beta F^{\mu\beta}.
\end{eqnarray}
The equation of motion (\ref{motion}) can be simply re-expressed in the
form
\begin{eqnarray}\label{motion2}
\partial_\mu F^{\mu\beta}=2\omega_\mu
F^{\mu\beta}-4\omega_c\Theta^{c\rho}(\partial_\rho
A_\mu)F^{\mu\beta}-2\omega_c\Theta^{c\rho} A_\mu\partial_\rho
F^{\mu\beta}.
\end{eqnarray}
Now using the fact that $\epsilon \omega=0$ yields the result. $\square$
\begin{remark}.

\begin{itemize}
\item The equation of motion
(\ref{motion2}) is reduced to $\partial_\mu F^{\mu\beta}=0$ in
ordinary Moyal plane.
\item The action of gauge theory covariantly
coupled  with the matter fields defined by
\begin{eqnarray}
\mathcal{S}=\mathcal{S}_{YM}+\mathcal{S}_M
\end{eqnarray}
where
\begin{eqnarray}
\mathcal{S}_{M}&=&\int_{\mathbb{R}^2}\,ed^2x\Big[\bar{\psi}(x)\Big(-i\Gamma^\mu
\nabla_\mu+m\Big)\psi(x)+\lambda_1(\bar{\psi}\star\psi\star\bar{\psi}\star\psi)(x)\cr
&&+\lambda_2(\bar{\psi}\star\bar{\psi}\star\psi\star\psi)(x)\Big]\star
e^{-1},
\end{eqnarray}
 is also invariant under global gauge transformation ($\delta\psi=i\alpha_0\psi,\,\,\,\delta\bar{\psi}=-i\alpha_0\bar{\psi}$).
 The current can be also easily deduced in the same manner as above.
 \end{itemize}
\end{remark}
\section{ Case of commuting vector fields}
Consider  the non coordinates base
$e_a^\mu=\delta_a^\mu+\omega_{ab}^\mu x^b$ and    the symmetric
tensor (between the index $a$ and $b$) $\omega_{ab}^{\mu}$. Then,
the twisted star product is naturally associative since
\begin{eqnarray}
[X_a,X_b]=\omega_{ba}^\mu\partial_{\mu}-\omega_{ab}^\mu\partial_{\mu}=0.
\end{eqnarray}
 The matrix representation  of  $e_a^\mu$ is given by
\begin{eqnarray}
(e)_a^\mu=\Big(\begin{array}{cc} 1+\omega_{11}^1x^1+\omega_{12}^1
x^2&\omega_{11}^2x^1+\omega_{12}^2x^2\\
\omega_{12}^1x^1+\omega_{22}^1x^2 &
1+\omega_{12}^2x^1+\omega_{22}^2 x^2
\end{array}\Big)
\end{eqnarray}
and
\begin{eqnarray}
(e)_\mu^a=\Big(\begin{array}{cc} 1-\omega_{11}^1x^1-\omega_{12}^1
x^2&-\omega_{11}^2x^1-\omega_{12}^2x^2\\
-\omega_{12}^1x^1-\omega_{22}^1x^2 &
1-\omega_{12}^2x^1-\omega_{22}^2 x^2
\end{array}\Big).
\end{eqnarray}
Further,
\begin{eqnarray}
e^{-1}&=&det(e_a^\mu)=1+(\omega_{11}^1+\omega_{12}^2)x^1+(\omega_{22}^2+\omega_{12}^1)x^2
\cr
e&=&det(e_\mu^a)=1-(\omega_{11}^1+\omega_{12}^2)x^1-(\omega_{22}^2+\omega_{12}^1)x^2.
\end{eqnarray}
The noncommutative tensor is provided by  $
(\widetilde{\Theta})^{\mu\nu}=\theta e^{-1}\Big(\begin{array}{cc}
0&1\\
-1&0
\end{array}\Big).
$ Besides, the matrix $e_{\mu}^a$ can be written as $
e_{\mu}^a=\delta_\mu^a+\omega^{ab}_\mu x_b,\,\,\, \mbox{ where
}\,\,\, \omega^{ab}_\mu=-\omega_{ab}^\mu .$ Finally, the solution
of the field equation $e_\mu^a=\partial_\mu\phi^a$ is well given
by  $ \phi^a=x^a+\frac{1}{2}\omega^{ab}_\mu x_b\,
x^\nu\delta_\nu^\mu $ as deduced in \cite{hd2}.  The $\phi^{c}$ variation of the
action can be easily computed and the resulting equation of motion
is
\begin{eqnarray}
\frac{\delta \mathcal{S}_{YM}}{\delta
\phi^{c}}=\mathcal{E}_{\phi^{c},A}=e^{-1}X_{a}(F_{\mu\nu}\star
F^{\mu\nu})-(X_{a}F^{\mu\nu})\{F_{\mu\nu},e^{-1}\}_{\star}=0
\end{eqnarray}
This variation generates the current
\begin{eqnarray}
\mathcal{K}^{\beta}&=&-\frac{ee_{b}^{\beta}}{4\kappa^2}\Big[(-F_{\mu\nu}\star
F^{\mu\nu}\star\delta\phi^{b}e^{-1})+T(\Delta)\Big(X_a(F_{\mu\nu}\star
F^{\mu\nu}),\widetilde{X}^{b}(\delta\phi^{a}e^{-1})\Big)\cr
&&-T(\Delta)\Big(\delta\phi^{a}(X_{a}F_{\mu\nu}),\widetilde{X}^{b}\{F^{\mu\nu},e^{-1}\}_{\star}
\Big)+\delta\phi^{b}(F_{\mu\nu}\star F^{\mu\nu}\star
e^{-1})\Big]\cr
&&+2S(\Delta)\Big(\delta\phi^{a}(X_{a}F_{\mu\nu})\star
e^{-1},\widetilde{X}^{b}F^{\mu\nu}\Big).
\end{eqnarray}
Performing the transformation $
\phi'^{c}(x)=\phi^{c}(x)+\delta\phi^{c}(x)$ where
$\delta\phi^{c}(x)=i\alpha \star \phi^{c}(x)$, with
$\alpha=\alpha_0$ or $\alpha=\alpha_0+\epsilon_\mu x^\mu$,  the variation of the
action yields the result:
\begin{eqnarray}
\delta\mathcal{S}_{YM}
&=&\int\,\mbox{d}^{2}x\,\,\delta\Big((\mathcal{L}_{YM}\star
e^{-1})e\Big)
\nonumber\\
&=&\int{\mbox{d}}^2
x\hspace{1mm}\Big\{\delta_{A_\mu}\Big((\mathcal{L}_{YM}\star
e^{-1})e\Big) +\delta_{\phi^{c}}\Big((\mathcal{L}_{YM}\star
e^{-1})e\Big)\Big\}\cr &=&\int_{M}{\mbox{d}}^2 x\hspace{1mm}
\partial_{\sigma}\Big(\mathcal{J}^{\sigma}+\mathcal{K}^{\sigma}\Big)=0
\end{eqnarray}
 Then $\mathcal{J}^{\sigma}$ can be
computed in the same way as for symmetric $\omega_{ab}^\mu$.  See
relation (\ref{cure}). The gauge invariance of the YM action
furnishes the current
$\mathcal{J}^{'\sigma}=\mathcal{J}^{\sigma}+\mathcal{K}^{\sigma}$.
Under  vanishing condition of the surface terms,
$\mathcal{J}^{'\sigma}$ is locally conserved on shell.

\section{Concluding remarks}
In this work, we have defined the twisted connections in
noncommutative spaces and discussed NC gauge transformations. Then, the
YM action, twisted in the dynamical Moyal space, has been proved to
be invariant under  $U_\star(1)$ gauge transformation with the parameter
$\alpha=\alpha_0+\epsilon_\mu x^\mu$, where $\epsilon_\mu$ is
an infinitesimal parameter and $\alpha_0$ a constant. 
   The gauge action is defined in 2 dimensional Moyal space
with signature $(1,1)$. The NC
gauge invariant currents are explicitly computed. These currents
are locally conserved.

Finally, it is worthy  mentioning that the approach developed here can be extended to investigate twisted
gauge theory in finite $D-$dimensional Moyal space. The only
technical difficulty resides in the fact that the choice of $\omega$ could not be arbitrarily made. For this
reason, the canonical form of $e_a^\mu$ given by
$\delta_a^\mu+\omega_{ab}^\mu x^b$ seems to be natural. The trivial case   $\omega=0$
 corresponds to NC YM theory, well known in the literature.

\section*{Acknowledgements}
This work is partially supported by the ICTP through the
OEA-ICMPA-Prj-15. The ICMPA is in partnership with the Daniel
Iagolnitzer foundation (DIF), France. The authors thank the referees and adjudicator for
their useful comments.

\section*{Appendix}
Notwithstanding  the condition  $[X_a,X_b]\neq 0,$ i.e.  $\omega_{ab}^\mu$ is skew-symmetric, the twisted
$\star-$product defined in (\ref{pppp}) remains noncommutative and associative. Indeed,
using the twisted star-product
 \begin{eqnarray}\label{prod}
(f\star g)(x)=m\Big[\exp\Big(\frac{i}{2}\theta
e^{-1}\epsilon^{\mu\nu}\partial_\mu\otimes\partial_\nu\Big)(f\otimes
g)(x)\Big],
\end{eqnarray}
 one can
see that
\begin{eqnarray}
e^{ikx}\star e^{iqx}=e^{i(k+q)x}e^{-\frac{i}{2}\theta
e^{-1}k\epsilon q}.
\end{eqnarray}
The Fourier transform of $f , g\in
\mathcal{S}(\mathbb{R}_{\tilde\Theta}^{2})$   can be written as
\begin{eqnarray}
\tilde{f}(k)=\int d^2x \,\,e^{-ikx}f(x),\,\,\,\,\tilde{g}(q)=\int
d^2x \,\,e^{-iqx}g(x)
\end{eqnarray}
with the function inverse transform given by
\begin{eqnarray}
f(x)=\int d^2k \,\,e^{ikx}\tilde{f}(k),\,\,\,\,g(x)=\int d^2q
\,\,e^{iqx}\tilde{g}(q).
\end{eqnarray}
We can  redefine the twisted  star-product of two Schwartz
functions $f , g$ as:
\begin{eqnarray}
(f\star g)(x)&=&\int
d^2kd^2q\,\,\tilde{f}(k)\tilde{g}(q)e^{ikx}\star e^{iqx}\cr
 &=& \int\,d^2kd^2q\,\,\tilde{f}(k)\tilde{g}(q)e^{i(k+q)x}e^{-\frac{i}{2}\theta
e^{-1}k\epsilon q}.
\end{eqnarray}
Then, we have
\begin{eqnarray}
 \Big((f\star g)\star
h\Big)(x)&=&\Big[\int\,d^2kd^2q\,\,\tilde{f}(k)\tilde{g}(q)e^{-\frac{i}{2}\theta
e^{-1}k\epsilon q}e^{i(k+q)x}\Big]\star\Big[\int d^2p
\,\,e^{ipx}\tilde{h}(p)\Big]\cr &=&\int\,d^2kd^2q
d^2p\,\,\tilde{f}(k)\tilde{g}(q)\tilde{h}(p)\Big(e^{-\frac{i}{2}\theta
e^{-1}k\epsilon q}e^{i(k+q)x}\Big)\star e^{ipx}
\end{eqnarray}
Recalling that $e^{-1}=1+\omega_\mu x^\mu,$ we get
\begin{eqnarray}\label{c1}
 \Big((f\star g)\star h\Big)(x)&=&\int\,d^2kd^2q
d^2p\,\,\tilde{f}(k)\tilde{g}(q)\tilde{h}(p)e^{-\frac{i}{2}\Big(\theta
k\epsilon q-\frac{1}{2}\theta^2\omega(k\epsilon q)\epsilon
p\Big)}e^{-\frac{i}{2}\theta e^{-1}(k+q)\epsilon p}\cr
&\times&e^{i\Big(k+q+p-\frac{1}{2}\theta\omega k\epsilon
q\Big)x}\cr &=&\int\,d^2kd^2q
d^2p\,\,\tilde{f}(k)\tilde{g}(q)\tilde{h}(p)e^{-\frac{i}{2}\Big(\theta
k\epsilon q+\theta(k+q)\epsilon
p-\frac{1}{2}\theta^2\omega(k\epsilon q)\epsilon p\Big)}\cr
&\times&e^{i\Big(k+q+p-\frac{1}{2}\theta \omega(k+q)\epsilon
p-\frac{1}{2}\theta\omega k\epsilon q\Big)x}
\end{eqnarray}
In the other side,
\begin{eqnarray}\label{c2}
\Big(f\star(g\star h)\Big)(x)&=&\int\,d^2kd^2q
d^2p\,\,\tilde{f}(k)\tilde{g}(q)\tilde{h}(p)e^{ikx}\star\Big(e^{-\frac{i}{2}\theta
e^{-1}q\epsilon p}e^{i(q+p)x}\Big)\cr &=& \int\,d^2kd^2q
d^2p\,\,\tilde{f}(k)\tilde{g}(q)\tilde{h}(p)e^{-\frac{i}{2}\Big(\theta
q\epsilon p+\theta k\epsilon(q+p)-\frac{1}{2}\theta^2\omega
(k\epsilon q)\epsilon p\Big)}\cr
&\times&e^{i\Big(k+q+p-\frac{1}{2}\theta\omega q\epsilon
p-\frac{1}{2}\theta\omega k\epsilon(q+p)\Big)x}
\end{eqnarray}
A straightforward expansion shows that  (\ref{c1}) and (\ref{c2}) 
 coincide. There results the conclusion that the used twisted star-product (\ref{pppp}) is well associative.

\end{document}